# A Massive Young Star-Forming Complex Study in Infrared and X-ray: X-ray Sources in Ten Star Forming Regions

Michael A. Kuhn, Konstantin V. Getman, Patrick S. Broos, Leisa K. Townsley, Eric D. Feigelson

*Dept. of Astronomy & Astrophysics, Pennsylvania State University, University Park PA 16802*

## ABSTRACT

The Massive Young star-forming complex Study in Infrared and X-ray (MYStIX) uses data from the *Chandra X-ray Observatory* to identify and characterize the young stellar populations in twenty Galactic ($d < 4$ kpc) massive star-forming regions. Here, the X-ray analysis for *Chandra* ACIS-I observations of ten of the MYStIX fields is described, and a catalog of >10,000 X-ray sources is presented. In comparison to other published *Chandra* source lists for the same regions, the number of MYStIX detected faint X-ray sources in a region is often doubled. While the higher catalog sensitivity increases the chance of false detections, it also increases the number of matches to infrared stars. X-ray emitting contaminants include foreground stars, background stars, and extragalactic sources. The X-ray properties of sources in these classes are discussed.

*Subject headings:* H II regions; stars: activity; stars: pre-main sequence; stars: formation; X-rays: stars

## 1. Introduction

The Massive Young star-forming complex Study in Infrared and X-ray (MYStIX) is a survey of twenty of the nearest ($d < 4$ kpc) massive star-forming regions (MSFRs) that have been observed with NASA's *Chandra X-ray Observatory* and with infrared (IR) survey telescopes (Feigelson et al. 2013). With ages less than ∼ 10 million years, each star-forming complex is dominated by OB stars and most have thousands of pre-main-sequence stars. Some portions of the stellar population are embedded in molecular clouds while for other portions the natal cloud material is ionized or dissipated.

X-ray observations are an effective strategy for obtaining a census of young stars in MSFRs. The X-ray luminosities of low-mass, pre-main-sequence stars are $10^2$-$10^5$ times greater than for main-sequence stars, and X-ray surveys are sensitive to both disk-bearing and disk-free, young, low-mass stars (see reviews by Güdel & Nazé 2009; Feigelson 2010). Massive OB stars also produce X-rays in shocks associated with their stellar winds (Lucy & White 1980; Owocki & Cohen 1999; Townsley et al. 2003). Both classes often exhibit a hard X-ray component that penetrates high column densities of interstellar material, and the sub-arcsecond on-axis resolution of *Chandra* probes the crowded centers of young stellar clusters.

The MYStIX project combines *Chandra* studies of MSFRs lying $0.4 - 3.6$ kpc from the Sun with near-IR surveys from the United Kingdom Infrared Telescope (UKIRT), often from the UKIDSS project (Lawrence et al. 2007), and mid-IR surveys from NASA's *Spitzer Space Telescope* Infrared Array Camera. Published OB stars are also included in the MYStIX samples. A full description of the MYStIX project is given by Feigelson et al. (2013). Together, MYStIX obtains large samples of the young stellar populations with a wide range of masses and ages for improved studies of star-cluster formation, cluster dynamical evolution, triggered star formation, and other issues relating to clustered star formation in giant molecular clouds.

This work describes the X-ray observations, data analysis, and resulting X-ray source lists and properties for ten MYStIX MSFRs. In order of increasing distance from the Sun (see Table 1 of Feigelson et al. 2013), they are: Flame Nebula, RCW 36, NGC 2264, Rosette Nebula, Lagoon Nebula, NGC 2362, DR 21, RCW 38, Trifid Nebula and NGC 1893. Additional MYStIX *Chandra* targets to be treated by Townsley & Broos (in preparation) are: NGC 6334, NGC 6357, Eagle Nebula, M 17, W 3, W 4, and NGC 3576. Previous studies have been published for much of the *Chandra* data; the MYStIX project reanalyzes these observations obtained from the *Chandra* archive in a unified fashion with methods that are tuned to finding weak sources in crowded environments with overlapping exposures. The MYStIX project adopts published X-ray source lists and properties for three MSFRs that were analyzed in a similar fashion: Orion Nebula (Getman et al. 2005); W 40 (Kuhn et al. 2010); and the Carina Nebula complex (Townsley et al. 2011; Broos et al. 2011a).

In this paper, Section 2 describes *Chandra* observations and data reduction, Section 3 describes the X-ray source lists, Section 4 compares our analysis to the results of prior X-ray studies of these regions, Section 5 discusses X-ray sources that are not young stars, Section 6 discusses distributions of X-ray flux and spectral hardness for various populations of X-ray sources, and Section 7 is the summary.

## 2. *Chandra* Observations and Data Reduction

X-ray observations were made with the imaging array on the Advanced CCD Imaging Spectrometer (ACIS-I; Garmire et al. 2003) on board the *Chandra X-ray Observatory* (CXO; Weisskopf et al. 2002). This array of four CCD detectors subtends $17' \times 17'$ on the sky. (We exclude data from the ACIS spectroscopic array due to *Chandra*'s reduced angular resolution far off axis.) The number of different *Chandra* pointings for each region, the total exposures for these pointings, and details of how the observations were taken are provided in Table 1. Data were acquired from the *Chandra* Data Archive[1] and were prepared from the "Level 1" data products derived from the satellite telemetry.

The *Chandra* observations were configured to meet the scientific goals of the original projects, so the spatial coverage and exposure durations vary among MSFRs. Sometimes a single ACIS pointing is available, while other targets have a mosaic of overlapping pointings. Furthermore, any pointing may be broken into several distinct observations, "ObsIDs," often due to spacecraft constraints that prevented longer continuous observations. Overall, 29 *Chandra* ObsIDs are included with typical integration times for a pointing of $40 - 100$ ks. Based on the X-ray luminosity function of the well-studied Orion Nebula Cluster (Feigelson et al. 2005) extrapolated to distances of $1 - 2$ kpc and a few visual magnitudes of line-of-sight absorption, these exposures are sufficient to detect most OB stars and lower mass pre-main-sequence stars down to ~0.5 - 1 $M_\odot$ for the MYStIX regions.

The ten MYStIX MSFRs treated here are listed in Table 2. Columns 1, 2, 3, and 5 are obtained from the MYStIX overview paper (Feigelson et al. 2013). For each *Chandra* mosaic, the areal coverage is given by Table 2, column 4. This is ~ 300 arcmin$^2$ for MSFRs with a single ACIS pointing: Flame Nebula, RCW 36, NGC 2362, DR 21, RCW 38, Trifid Nebula, and NGC 1893. The Lagoon Nebula has two overlapping pointings while NGC 2264 has three pointings. The Rosette Nebula is a linear mosaic of six pointings, deeper on the central cluster NGC 2244 and shallower through the Rosette Molecular Cloud. Columns $6 - 9$ give the number of X-ray sources obtained from the procedures outlined below. Figure 1 shows the X-ray mosaics in which hundreds to thousands of young stars are seen.

---

[1] http://cxc.harvard.edu/cda

Table 1. Log of *Chandra* Observations

| ObsID | Sequence | Start Time (UT) | Exposure (s) | Aimpoint $\alpha_{J2000}$ | $\delta_{J2000}$ | Roll Angle (°) | Mode | PI |
|---|---|---|---|---|---|---|---|---|
| **Flame Nebula** | | | | | | | | |
| 1878 | 200106 | 2001-08-08T06:37 | 75457 | 05:41:46.30 | −01:55:28.7 | 111 | Faint | S. Skinner |
| **RCW 36** | | | | | | | | |
| 6433 | 200407 | 2006-09-23T19:44 | 69708 | 08:59:27.49 | −43:45:27.0 | 124 | Very Faint | M. Tsujimoto |
| **NGC 2264** | | | | | | | | |
| 2540 | 200148 | 2002-10-28T14:46 | 95142 | 06:40:58.09 | +09:34:00.4 | 78 | Faint | S. Sciortino |
| 2550 | 200158 | 2002-02-09T05:10 | 48134 | 06:40:48.00 | +09:50:59.9 | 281 | Faint | J. Stauffer |
| 9768 | 200526 | 2008-03-12T17:55 | 27786 | 06:41:11.99 | +09:30:00.0 | 270 | Faint | G. Micela |
| 9769 | 200526 | 2008-03-28T14:47 | 29756 | 06:41:11.99 | +09:30:00.0 | 266 | Faint | G. Micela |
| **Rosette Nebula** | | | | | | | | |
| 1874 | 200102 | 2001-01-05T11:54 | 15950 | 06:31:52.00 | +04:55:57.0 | 335 | Faint | L. Townsley |
| 1875 | 200103 | 2001-01-05T17:47 | 19503 | 06:32:40.00 | +04:43:00.0 | 335 | Faint | L. Townsley |
| 1876 | 200104 | 2001-01-05T23:29 | 19506 | 06:33:16.30 | +04:34:56.9 | 335 | Faint | L. Townsley |
| 1877 | 200105 | 2001-01-06T05:11 | 19506 | 06:34:16.49 | +04:28:00.9 | 335 | Faint | L. Townsley |
| 1877 | 200220 | 2004-01-01T02:20 | 74999 | 06:31:55.50 | +04:56:34.0 | 351 | Very Faint | L. Townsley |
| 8454 | 200459 | 2007-02-09T02:00 | 11844 | 06:30:50.39 | +04:59:34.0 | 286 | Very Faint | G. Garmire |
| 12142 | 200657 | 2010-12-10T04:36 | 39540 | 06:34:37.60 | +04:12:44.2 | 41 | Very Faint | G. Garmire |
| **Lagoon Nebula** | | | | | | | | |
| 977 | 200084 | 2001-06-18T11:40 | 59598 | 18:04:24.00 | −24:21:20.0 | 80 | Faint | S. Murray |
| 3754 | 200224 | 2003-07-25T17:28 | 108512 | 18:03:45.10 | −24:22:05.0 | 272 | Very Faint | M. Gagné |
| 4297 | 200224 | 2003-07-24T10:08 | 14634 | 18:03:45.10 | −24:22:05.0 | 272 | Very Faint | M. Gagné |
| 4444 | 200224 | 2003-07-28T00:01 | 29806 | 18:03:45.10 | −24:22:05.0 | 272 | Very Faint | M. Gagné |
| **NGC 2362** | | | | | | | | |
| 4469 | 200238 | 2003-12-23T03:03 | 97883 | 07:18:42.79 | −24:57:18.5 | 22 | Very Faint | S. Murray |
| **DR 21** | | | | | | | | |
| 7444 | 200454 | 2007-08-22T03:28 | 48706 | 20:39:00.70 | +42:18:56.8 | 204 | Very Faint | F. Damiani |
| 8598 | 200454 | 2007-11-27T23:40 | 20480 | 20:39:00.70 | +42:18:56.8 | 303 | Very Faint | F. Damiani |
| 9770 | 200454 | 2007-11-29T15:12 | 18986 | 20:39:00.70 | +42:18:56.8 | 303 | Very Faint | F. Damiani |
| 9771 | 200454 | 2007-12-02T09:50 | 9143 | 20:39:00.70 | +42:18:56.8 | 303 | Very Faint | F. Damiani |
| **RCW 38** | | | | | | | | |
| 2556 | 200164 | 2001-12-10T10:14 | 95924 | 08:59:19.20 | −47:30:21.9 | 51 | Very Faint | S. Wolk |
| **Trifid Nebula** | | | | | | | | |
| 2566 | 200174 | 2002-06-13T02:17 | 58061 | 18:02:30.30 | −23:01:29.3 | 105 | Faint | J. Rho |
| **NGC 1893** | | | | | | | | |
| 6406 | 200383 | 2006-11-09T12:51 | 115662 | 05:22:49.99 | +33:28:05.0 | 107 | Faint | G. Micela |
| 6407 | 200383 | 2006-11-15T05:31 | 126217 | 05:22:49.99 | +33:28:05.0 | 107 | Faint | G. Micela |
| 6408 | 200383 | 2007-01-23T00:12 | 102828 | 05:22:49.99 | +33:28:05.0 | 262 | Faint | G. Micela |
| 8462 | 200383 | 2006-11-07T13:33 | 42615 | 05:22:49.99 | +33:28:05.0 | 107 | Faint | G. Micela |
| 8476 | 200383 | 2006-11-17T10:55 | 53268 | 05:22:49.99 | +33:28:05.0 | 107 | Faint | G. Micela |

Note. — Exposure times are the net usable times after various filtering steps are applied in the data reduction process. The aimpoints and roll angles are obtained from the satellite aspect solution before astrometric correction is applied. Units of right ascension are hours, minutes, and seconds; units of declination are degrees, arcminutes, and arcseconds.

Table 2. X-ray Observations of Ten MYStIX Star Forming Regions

| Name | Location ($\alpha, \delta$) | Distance kpc | Area deg$^2$ | Area pc$^2$ | Previous publs | Number of X-ray sources[a] Tot | Faint | Mod | Strong |
|---|---|---|---|---|---|---|---|---|---|
| (1) | (2) | (3) | (4) | | (5) | (6) | (7) | (8) | (9) |
| Flame Nebula | 0541−01 | 0.414 | 0.08 | (4) | 1, 2 | 547 | 278 | 167 | 102 |
| RCW 36 | 0859−43 | 0.7±0.2 | 0.08 | (12) | ... | 502 | 296 | 175 | 31 |
| NGC 2264 | 0641+09 | 0.913±0.1 | 0.19 | (52) | 3, 4, 5, 6 | 1,328 | 590 | 479 | 259 |
| Rosette Nebula | 0632+05 | 1.33±0.05 | 0.46 | (248) | 7, 8, 9, 10 | 1,962 | 1,153 | 709 | 100 |
| Lagoon Nebula | 1804−24 | $1.3^{+0.5}$ | 0.13 | (69) | 11, 12 | 2,427 | 981 | 1,132 | 314 |
| NGC 2362 | 0718−25 | 1.48 | 0.08 | (55) | 13, 14, 15 | 690 | 324 | 302 | 64 |
| DR 21 | 2039+42 | 1.50±0.08 | 0.09 | (62) | ... | 765 | 431 | 312 | 22 |
| RCW 38 | 0859−47 | 1.7±0.9 | 0.08 | (71) | 16, 17, 18 | 1,019 | 517 | 436 | 66 |
| Trifid Nebula | 1802−23 | 2.7±0.5 | 0.08 | (182) | 19 | 633 | 344 | 254 | 35 |
| NGC 1893 | 0523+33 | 3.6±0.2 | 0.10 | (379) | 20, 21 | 1,442 | 447 | 786 | 209 |

[a]Tot = total number of *Chandra* sources. "Faint" sources have < 10 counts, "Mod" sources have 10 − 100 counts, "Strong" sources have > 100 counts. These are net extracted counts after background subtraction in the *Chandra* total band (0.5-8 keV).

Note. — X-ray publications: 1. Skinner et al. (2003) 2. Ezoe et al. (2006) 3. Ramírez et al. (2004) 4. Sung et al. (2004) 5. Rebull et al. (2006) 6. Flaccomio et al. (2006) 7. Townsley et al. (2003) 8. Wang et al. (2008) 9. Wang et al. (2009) 10. Wang et al. (2010) 11. Damiani et al. (2004) 12. Henderson & Stassun (2012) 13. Delgado et al. (2006) 14. Damiani et al. (2006) 15. Dahm et al. (2007) 16. Wolk et al. (2002) 17. Wolk et al. (2006) 18. Winston et al. (2011) 19. Rho et al. (2004) 20. Caramazza et al. (2008) 21. Caramazza et al. (2012)

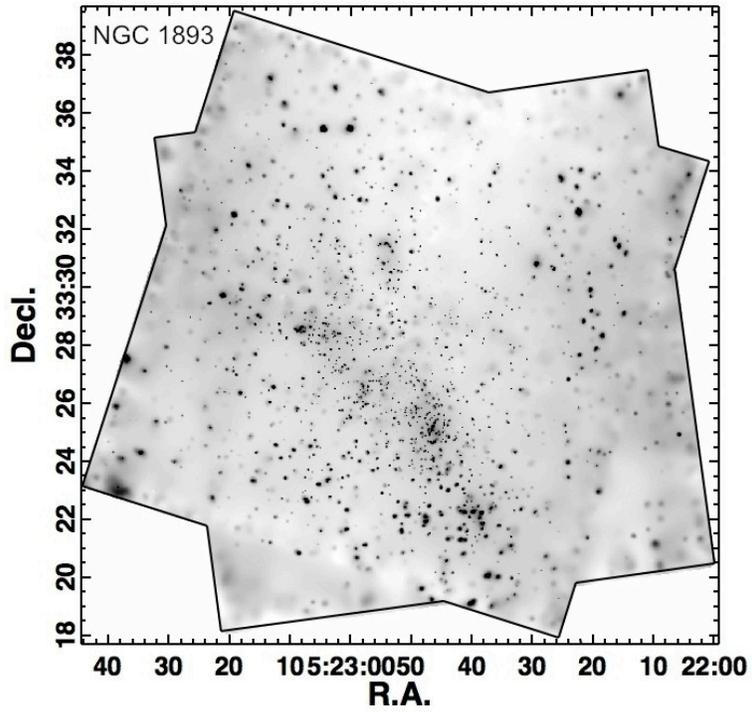

Fig. 1.— Adaptively smoothed ACIS-I mosaic images of the NGC 1893 MSFRs shown in logarithmic scale. Smoothing has been performed on the total $0.5 - 8.0$ keV band. Corresponding figures for each MSFR listed in Table 2 are available in the electronic edition of this article.

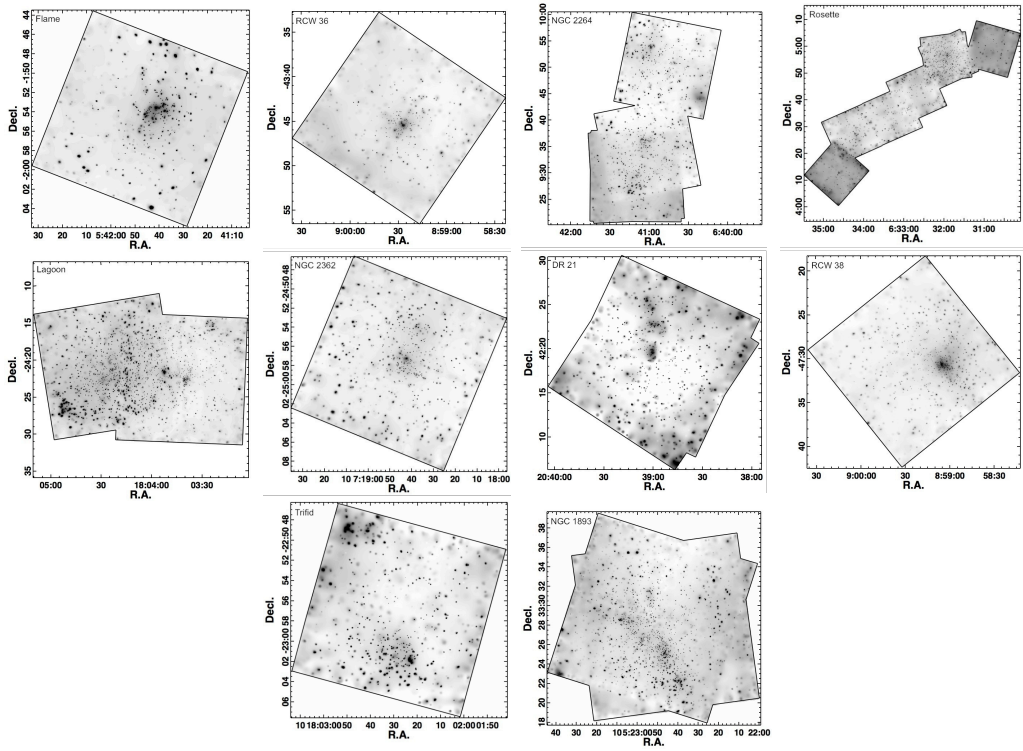

Fig. 1.— electronic figure set

The methodology of the MYStIX data reduction follows closely the procedures established by the *Chandra* Carina Complex Project (CCCP; Townsley et al. 2011). These procedures are described in Broos et al. (2010), Getman et al. (2010), and Broos et al. (2011a). The analysis uses codes from CIAO (Fruscione et al. 2006), MARX (Davis et al. 2012), HEASoft[2], and the Astronomy User's Library (Landsman 1993), which are integrated by scripts in the IDL language. The analysis combines all imaging-mode ACIS-I data that is available from the *Chandra* archive to create improved *Chandra* point-source catalogs for the regions in our study. Compared to procedures used by many previous researchers, the methodological improvements allow up to twice as many X-ray point sources to be detected (Broos et al. 2011a). We briefly summarize the data reduction methodology, including updates to the Carina analysis procedures that are outlined here and presented in detail by Townsley & Broos (in preparation).

Following Broos et al. (2010), the "Level 2" data products are rebuilt from the "Level 1" products. This processing involves event energy calibration, improvements to the position of individual events, and cleaning of contamination by bad pixels, background flaring, and cosmic-ray afterglows. Heavily cleaned event lists are used for source detection and extraction of weak sources, while lightly cleaned data are used for extraction of bright sources. Alignment of the ObsID coordinate systems is adjusted using the positions of sources from the Two Micron All Sky Survey (2MASS; Skrutskie et al. 2006), and multiple ObsIDs are tiled to create a *Chandra* mosaic image.

The most common source-detection procedures for *Chandra* imaging observations of star-forming regions are based on the wavelet transformation using the *wavdetect* procedure developed by Freeman et al. (2002) or the *PWDetect* procedure developed by Damiani et al. (1997). These methods do not use calibration files that describe the complex shapes and strong spatial variations in the telescope point-spread function due to the unusual optics needed for wide-field X-ray imaging, so they are not optimal at detecting the faintest sources, resolving closely spaced sources in a crowded field, or recovering source positions for far off-axis sources. We use instead a maximum likelihood reconstruction applied to X-ray events in small tiles across the field that accurately traces spatial variations in the point-spread function, a procedure developed by Townsley et al. (2006). The reconstruction is calculated with the Lucy-Richardson algorithm (Lucy 1974) that is an implementation of the EM Algorithm widely used in maximum likelihood statistical calculations (McLachlan & Krishnan 2008). A bump-hunting algorithm is then applied to the reconstructed image to give a superset of candidate point sources.

---

[2]https://heasarc.gsfc.nasa.gov/lheasoft

The *ACIS Extract*[3] package (Broos et al. 2010, 2012) is then applied to the original data to extract events for each of the candidate point sources. This is an elaborate procedure that again uses the local point-spread function to estimate source position, create an extraction region that does not overlap neighboring extraction regions, measure a background from an optimized region that accounts for contamination by neighboring sources, and compute the null probability, $ProbNoSrc\_min$, that no source exists given the observed local background level, assuming the events follow the Poisson distribution. Starting with the candidate point sources, insignificant sources are pruned and source positions, extraction regions, backgrounds, and $ProbNoSrc\_min$ values are recomputed in an iterative process until convergence of the source list is achieved. Candidate sources that satisfy $ProbNoSrc\_min < 0.01$, or a 99% probability that the source is not a random fluctuation in the background, are considered to be valid. This criterion can sometimes detect on-axis sources down to 3 extracted counts, considerably fainter than most wavelet methods. A fraction of the faintest X-ray sources may thus be spurious. We choose to produce the most sensitive possible X-ray catalog because, in our experience with MYStIX and earlier studies, even the faintest X-ray sources are often associated with IR-detected young stars. The X-ray positions typically have subarcsecond accuracy; sources within $3'$ of the field center have median 1-$\sigma$ uncertainties around $0.15''$, increasing to $\sim 0.8''$ at $10'$ off-axis. Thus, spurious X-ray sources will rarely be incorrectly matched with IR sources. Our highly sensitive X-ray catalog plays a major role in producing rich MYStIX catalogs of probable members of the observed star-forming complexes (Broos et al. 2013).

These methods are more fully presented in Broos et al. (2011a) where they were used for the identification of $> 14,000$ X-ray sources in the CCCP. The MYStIX analysis is somewhat improved over the CCCP analysis. First, astrometric alignment is improved. Second, candidate sources within an arcsecond of a bright source are visually evaluated as to whether they are possible examples of an artifact found in the Chandra point-spread function[4] and removed if consistent with that artifact. Third, potential background regions now account for CCD photon pile-up and readout trails associated with bright sources. These improvements are described by Townsley & Broos (in preparation).

Source properties such as position, flux, spectral hardness, and variability are then computed with methods adapted to the low counts seen in most MYStIX fields (see columns $7-9$ of Table 2). As many MYStIX sources are too weak for statistical fitting of spectral models, nonparametric procedures to estimate hydrogen column densities and absorption-corrected fluxes described by Getman et al. (XPHOT; 2010) are needed[5]. An important element is the use of the median energy of extracted events, the most robust estimate of spectral hardness, as an estimator of interstellar absorption. This is combined with prior empirical knowledge (e.g., from the Chandra Orion Ultradeep Project; Getman et al. 2005) of young stellar X-ray spectra to obtain good estimates of absorption-corrected source fluxes and their systematic and statistical errors. This analysis assumes the X-ray spectral shapes of young, low-mass stars. Therefore, XPHOT quantities will be unreliable for high-mass stars that are dominated by X-ray emission from a stellar wind.

---

[3]The *ACIS Extract* software package and User's Guide are available at http://www.astro.psu.edu/xray/acis/acis_analysis.html.

[4]http://cxc.harvard.edu/ciao4.4/caveats/psf_artifact.html

## 3. List of X-ray Sources

A total of 11,315 X-ray sources were identified and extracted. Columns 6 − 9 of Table 2 summarize the resulting source populations in the ten MYStIX fields analyzed here. Most of the X-ray sources are quite faint with typical sources exhibiting around 10 counts, although source counts range from ≤4 to >1000. Bright sources are occasionally affected by pile-up, which will bias inferred X-ray properties, decreasing X-ray count rates and increasing spectral hardness. Sources with significant pile-up are listed in Table 3.

Table 4 lists X-ray properties obtained from ACIS Extract and XPHOT. Photometry is computed for three bands, soft (0.5 − 2.0 keV), hard (2.0 − 8.0 keV), and total (0.5 − 8.0 keV). Descriptions of how ACIS Extract quantities are calculated are given by Broos et al. (2010) and Broos et al. (2011a). Of particular interest are median energy and photon flux values, which may be directly compared between different *Chandra* ACIS-I observations. Inferred spectroscopic properties from XPHOT include hydrogen column density, X-ray flux (both incident and absorption corrected) in the hard and total bands, and systematic and statistical errors on these quantities.

Point-source detection sensitivity is related to photon flux in the bands used for source detection[6]. However, sensitivity varies across the fields of view due to telescope vignetting, degradation of the point-spread function with off-axis angle, and overlapping exposures of different durations in mosaics. In particular, the "egg-crate effect" was identified by Broos et al. (2011a) where there is a higher density of X-ray sources near the focal axis of the telescope due to increased sensitivity. Similar effects are seen in MYStIX. For example, there are a number of weak, ≤4 count sources near the centers of the ACIS-I pointings, many of which have hard X-ray spectra and no IR counterpart – consistent with being extragalactic X-ray sources and/or spurious detections. Some MYStIX science studies require uniform-in-flux source samples across the entire regions; for example, the spatial structure study of Kuhn et al. (2013) uses the empirical distributions of X-ray sources at various off-axis angles to calculate the incident $0.5 - 8.0$ keV photon flux above which the sample is complete.

---

[5]The XPHOT software package is available from http://www2.astro.psu.edu/users/gkosta/XPHOT/. See also Getman et al. (2012).

[6]Detection was performed in three energy ranges, 0.5 − 2.0, 0.5 − 7.0, and 2.0 − 7.0 keV, that differ from the photometry ranges above. To achieve greatest sensitivity, we exclude events with energies above 7 keV where the background levels increase.

Table 3. X-Ray Sources Affected by Pile-up

| Region (1) | IAU source name (2) |
|---|---|
| Flame | 054137.74-015351.5 |
|  | 054138.24-015309.1 |
|  | 054138.58-015322.8 |
|  | 054141.35-015332.7 |
|  | 054146.11-015414.8 |
|  | 054148.21-015601.9 |
| NGC 2264 | 064040.44+095050.4 |
|  | 064046.07+094917.3 |
|  | 064046.07+094917.3 |
|  | 064056.50+095410.5 |
|  | 064058.50+093331.7 |
|  | 064058.66+095344.8 |
|  | 064103.49+093118.4 |
|  | 064105.36+093313.5 |
|  | 064105.54+093140.6 |
|  | 064105.54+093140.6 |
|  | 064105.74+093101.3 |
|  | 064106.19+093622.9 |
|  | 064106.82+092732.2 |
|  | 064108.59+092933.7 |
|  | 064110.00+092746.0 |
|  | 064113.03+092732.0 |
| Rosette | 063155.51+045634.3 |
|  | 063231.43+044234.0 |
|  | 063327.50+043557.2 |
| Lagoon | 180352.45-242138.5 |
|  | 180407.36-242221.9 |
|  | 180414.63-242155.6 |
| NGC 2362 | 071842.48-245715.8 |
|  | 071845.25-245643.9 |
|  | 071850.46-245754.7 |
| RCW 38 | 085905.65-473040.9 |

The determination of whether an X-ray source is a young star in the target MSFR is not made here. MYStIX Probable Complex Members (MPCMs) are classified using X-ray and IR properties by Broos et al. (2013) using the "Naive Bayes Classifier," a supervised machine-learning algorithm. X-ray properties used by this algorithm include X-ray median energy and X-ray variability. We examine some of the X-ray properties of these classes in §6. In the MYStIX fields, typically 65% to 80% of X-ray sources are designated MPCMs (Broos et al. 2013, their Table 6).

Table 4.  MYStIX X-ray Sources and Properties

| Column Label | | | |
|---|---|---|---|
| Electronic (1) | In-print (2) | Units (3) | Description (4) |

**X-ray Photometry** (Broos et al. 2010, ACIS Extract)

| | | | |
|---|---|---|---|
| TBD | MYSTIX_SFR | ⋯ | name of MYStIX star-forming region |
| Name | Name | ⋯ | IAU source name; prefix is CXO J (*Chandra X-ray Observatory*) |
| Label | Label[a] | ⋯ | source name used within the project |
| RAdeg | RAdeg | deg | right ascension (J2000) |
| DEdeg | DEdeg | deg | declination (J2000) |
| PosErr | PosErr | arcsec | 1-$\sigma$ error circle around (RAdeg,DEdeg) |
| PosType | PosType | ⋯ | algorithm used to estimate position (Broos et al. 2010, §7.1) |
| PNoSrc-m | ProbNoSrc_min | ⋯ | smallest of ProbNoSrc_t, ProbNoSrc_s, ProbNoSrc_h |
| PNoSrc-t | ProbNoSrc_t | ⋯ | p-value[b] for no-source hypothesis (Broos et al. 2010, §4.3) |
| PNoSrc-s | ProbNoSrc_s | ⋯ | p-value for no-source hypothesis |
| PNoSrc-h | ProbNoSrc_h | ⋯ | p-value for no-source hypothesis |
| PKS-s | ProbKS_single[c] | ⋯ | smallest p-value for the one-sample Kolmogorov-Smirnov statistic under the no-variability null hypothesis within a single-observation |
| PKS-m | ProbKS_merge[c] | ⋯ | smallest p-value for the one-sample Kolmogorov-Smirnov statistic under the no-variability null hypothesis over merged observations |
| ExpNom | ExposureTimeNominal | s | total exposure time in merged observations |
| ExpFrac | ExposureFraction[d] | ⋯ | fraction of ExposureTimeNominal that source was observed |
| NObs | NumObservations | ⋯ | total number of observations extracted |
| NMerge | NumMerged | ⋯ | number of observations merged to estimate photometry properties |
| MergeB | MergeBias | ⋯ | fraction of exposure discarded in merge |
| e_Theta | Theta_Lo | arcmin | smallest off-axis angle for merged observations |
| Theta | Theta | arcmin | average off-axis angle for merged observations |
| E_Theta | Theta_Hi | arcmin | largest off-axis angle for merged observations |
| PSFFrac | PsfFraction | ⋯ | average PSF fraction (at 1.5 keV) for merged observations |
| SrcArea | SrcArea | (0.492 arcsec)$^2$ | average aperture area for merged observations |
| AGlow | AfterglowFraction[e] | ⋯ | suspected afterglow fraction |
| SrCnt-t | SrcCounts_t | count | observed counts in merged apertures |
| SrCnt-s | SrcCounts_s | count | observed counts in merged apertures |
| SrCnt-h | SrcCounts_h | count | observed counts in merged apertures |
| BkgScl | BkgScaling | ⋯ | scaling of the background extraction (Broos et al. 2010, §5.4) |
| BkgCt-t | BkgCounts_t | count | observed counts in merged background regions |
| BkgCt-s | BkgCounts_s | count | observed counts in merged background regions |
| BkgCt-h | BkgCounts_h | count | observed counts in merged background regions |

Table 4—Continued

| Column Label | | Units | Description |
|---|---|---|---|
| Electronic (1) | In-print (2) | (3) | (4) |
| NCt-t | NetCounts_t | count | net counts in merged apertures |
| NCt-s | NetCounts_s | count | net counts in merged apertures |
| NCt-h | NetCounts_h | count | net counts in merged apertures |
| loNCt-t | NetCounts_Lo_t[f] | count | 1-sigma lower bound on NetCounts_t |
| upNCt-t | NetCounts_Hi_t | count | 1-sigma upper bound on NetCounts_t |
| loNCt-s | NetCounts_Lo_s | count | 1-sigma lower bound on NetCounts_s |
| upNCt-s | NetCounts_Hi_s | count | 1-sigma upper bound on NetCounts_s |
| loNCt-h | NetCounts_Lo_h | count | 1-sigma lower bound on NetCounts_h |
| upNCt-h | NetCounts_Hi_h | count | 1-sigma upper bound on NetCounts_h |
| Area-t | MeanEffectiveArea_t[g] | $cm^2$ count $photon^{-1}$ | mean ARF value |
| Area-s | MeanEffectiveArea_s | $cm^2$ count $photon^{-1}$ | mean ARF value |
| Area-h | MeanEffectiveArea_h | $cm^2$ count $photon^{-1}$ | mean ARF value |
| Eng-t | MedianEnergy_t[h] | keV | median energy, observed spectrum |
| Eng-s | MedianEnergy_s | keV | median energy, observed spectrum |
| Eng-h | MedianEnergy_h | keV | median energy, observed spectrum |
| TBD | PhotonFlux_t[i] | photon /$cm^2$ /s | log incident photon flux |
| TBD | PhotonFlux_s | photon /$cm^2$ /s | log incident photon flux |
| TBD | PhotonFlux_h | photon /$cm^2$ /s | log incident photon flux |
| **X-ray Spectral Model** (Getman et al. 2010, XPHOT) | | | |
| TBD | (F_H)[j] | erg /$cm^2$ /s | X-ray flux, 2:8 keV |
| TBD | (F_HC) | erg /$cm^2$ /s | absorption-corrected X-ray flux, 2:8 keV |
| TBD | (SF_HC_STAT) | erg /$cm^2$ /s | 1-sigma statistical uncertainty on FX_HC |
| TBD | (SF_HC_SYST) | erg /$cm^2$ /s | 1-sigma systematic uncertainty on FX_HC |
| TBD | F_T | erg /$cm^2$ /s | X-ray flux, 0.5:8 keV |
| logL | (F_TC) | erg /$cm^2$ /s | absorption-corrected X-ray flux, 0.5:8 kev |
| TBD | (SF_TC_STAT) | erg /$cm^2$ /s | 1-sigma statistical uncertainty on FX_TC |
| TBD | (SF_TC_SYST) | erg /$cm^2$ /s | 1-sigma systematic uncertainty on FX_TC |
| logNH | (LOGNH_OUT) | /$cm^2$ | gas column density |
| TBD | (SLOGNH_OUT_STAT_OUT) | /$cm^2$ | 1-sigma statistical uncertainty on LOGNH_OUT |
| TBD | (SLOGNH_OUT_SYST_OUT) | /$cm^2$ | 1-sigma systematic uncertainty on LOGNH_OUT |

Note. — Rows are sorted by R.A.

Note. — Col. (1): Column label chosen by ApJ and used in the electronic edition of this article.

Col. (2): Column label previously published by the CCCP (Broos et al. 2011a) and produced by the ACIS Extract software.

Note. — The suffixes "_t", "_s", and "_h" on names of photometric quantities designate the *total* (0.5–8 keV), *soft* (0.5–2 keV), and *hard* (2–8 keV) energy bands.

Note. — Source significance quantities (ProbNoSrc_t, ProbNoSrc_s, ProbNoSrc_h, ProbNoSrc_min) are computed using a subset of each source's extractions chosen to maximize significance (Broos et al. 2010, §6.2). Source position quantities (RAdeg, DEdeg, PosErr) are computed using a subset of each source's extractions chosen to minimize the position uncertainty (Broos et al. 2010, §6.2 and 7.1). All other quantities are computed using a subset of each source's extractions chosen to balance the conflicting goals of minimizing photometric uncertainty and of avoiding photometric bias (Broos et al. 2010, §6.2 and 7).

[a]Source labels identify a *Chandra* pointing; they do not convey membership in astrophysical clusters.

[b]In statistical hypothesis testing, the p-value is the probability of obtaining a test statistic at least as extreme as the one that was actually observed when the null hypothesis is true.

[c]See Broos et al. (2010, §7.6) for a description of the variability metrics, and caveats regarding possible spurious indications of variability using the ProbKS_merge metric.

[d]Due to dithering over inactive portions of the focal plane, a *Chandra* source is often not observed during some fraction of the nominal exposure time. (See http://cxc.harvard.edu/ciao/why/dither.html.) The reported quantity is FRACEXPO produced by the *CIAO* tool *mkarf*.

[e]Some background events arising from an effect known as "afterglow" (http://cxc.harvard.edu/ciao/why/afterglow.html) may contaminate source extractions, despite careful procedures to identify and remove them during data preparation (Broos et al. 2010, §3). After extraction, we attempt to identify afterglow events using the tool ae_afterglow_report, and report the fraction of extracted events attributed to afterglow; see the *ACIS Extract* manual (http://www.astro.psu.edu/xray/acis/acis_analysis.html).

[f]Confidence intervals (68%) for NetCounts quantities are estimated by the *CIAO* tool *aprates* (http://asc.harvard.edu/ciao/ahelp/aprates.html).

[g]The ancillary response file (ARF) in ACIS data analysis represents both the effective area of the observatory and the fraction of the observation for which data were actually collected for the source (ExposureFraction).

[h]MedianEnergy is the median energy of extracted events, corrected for background (Broos et al. 2010, §7.3).

[i]PhotonFlux = (NetCounts / MeanEffectiveArea / ExposureTimeNominal) (Broos et al. 2010, §7.4)

[j]XPHOT assumes X-ray spectral shapes of young, low-mass stars, which come from coronal X-ray emission. XPHOT quantities will therefore be unreliable for high-mass stars, for which X-ray emission is associated with the stellar wind (Getman et al. 2010).

## 4. Comparison to Published Source Catalogs

As indicated in Table 2 (column 5), most of these *Chandra* fields have been previously analyzed. Generally, the MYStIX analysis emerges with considerably larger source lists than previous studies. This is partly due to improved methods for super-resolving crowded regions using maximum likelihood image reconstruction and improved treatment of local background levels. But the MYStIX procedure also sets a lower threshold for source existence ($ProbNoSrc\_min = 1\%$) compared to many other analyses. MYStIX will thus have more false positive (spurious) sources near the detection threshold[7]. This was purposefully done because the lower threshold also picks up many real faint sources. Possibly spurious sources can be excluded from the MPCM sample of probable members because they will not have necessary properties (e.g., match to infrared source, X-ray flaring) for classification by Broos et al. (2013). These sources are generally ranked as "unclassified." However, past experience indicates that many of the faintest sources have IR counterparts, and improved near-IR data has often lead to new IR counterparts to the faintest ACIS Extract X-ray sources that had previously been "unclassified" (e.g., Broos et al. 2011b). This is expected because the pre-main sequence X-ray luminosity function rises steeply around $\log L_X \sim 30$ erg/s where $ProbNoSrc\_min \sim 0.1 - 1\%$ in most MYStIX fields.

Global comparisons with previous work can be directly made for some MYStIX MSFRs. Other cases used different datasets in past studies, or used data analysis procedures similar to those in MYStIX.

**Flame Nebula** Skinner et al. (2003) analyzed this early *Chandra* field using standard procedures recommended by the *Chandra X-ray Center*. The *wavdetect* wavelet source-detection algorithm (Freeman et al. 2002) was used with different scale factors and thresholds, followed by visual inspection to remove spurious detections and add missed sources. They emerged with 283 X-ray sources, of which 248 were spatially associated with IR stars or radio sources indicating likely memberships in the Orion/Flame young stellar population. Our MYStIX analysis of this dataset obtained 547 X-ray sources, of which 422 are identified later as MPCMs (Broos et al. 2013).

---

[7] Monte Carlo simulation of source detection with artificial sources is a common procedure for evaluating a detection algorithm; however, several reasons why this procedure will not produce useful information in this case are described by Broos et al. (2011a, Section 6.2). Briefly, the complexities of the *Chandra* survey and detection procedures limit the accuracy of simulations. Furthermore, propagation of simulated spurious sources through additional sample selection (e.g. X-ray/IR matching, source classification) and science analysis would be difficult. Broos et al. (2011a, their figure 9) show the number of X-ray sources with and without near-IR counterparts stratified by $ProbNoSrc\_min$, where the number of X-ray sources without counterparts is an upper limit to the number of spurious sources. The fraction without counterparts increases with larger $ProbNoSrc\_min$, but even near the $ProbNoSrc\_min = 0.01$ detection threshold there is no sudden increase in this category. Similar trends are found in MYStIX.

**NGC 2362** Damiani et al. (2006) obtained 387 sources using the *PWDetect* wavelet source-detection algorithm (Damiani et al. 1997) with a conservative threshold set for < 1 spurious detection in the *Chandra* field of view. This is a single *Chandra* exposure without significant crowding, and thus does not present difficulties to traditional source detections. MYStIX obtained 690 sources, mainly due to choice of a lower threshold for faint source existence, of which 467 are classified as MPCMs.

**RCW 38** Wolk et al. (2006) and Winston et al. (2011) used *PWDetect* and *wavdetect* source-detection algorithms with a nominal false detection rate of <1% to find 518 sources. In a separate analysis, Evans et al. (2010) found 410 sources with ≥ 10 net counts using the *Chandra Source Catalog* algorithm. Our MYStIX analysis gives 1019 sources, nearly twice the number found by Winston et al. (2011) from the same dataset, of which 813 are classified as MPCMs.

**Trifid Nebula** Rho et al. (2004) made several runs of *wavdetect* with different wavelet scales on this image and a conservative faint source threshold to obtain 353 *Chandra* sources. The MYStIX analysis gives 633 sources, of which 418 are classified as MPCMs.

**NGC 1893** Application of *PWDetect* with a threshold to give ∼ 10 spurious sources resulted in 1025 *Chandra* sources (Caramazza et al. 2008, 2012). MYStIX analysis produced 1442 sources, of which 1110 are classified as MPCMs.

Here we make three detailed comparisons between previous *Chandra* source detections and the MYStIX procedures to further understanding of the source detection capabilities.

**Trifid Nebula** Figure 2 shows the core of the cluster ionizing the Trifid Nebula imaged by *Chandra* and UKIRT. The 60 ks *Chandra* ACIS-I exposure is among the lowest for MYStIX while this is one of the most distant regions at 2.7 kpc. The center of the cluster is off axis in the *Chandra* image resulting in an elongated point-spread function. This region is projected near the Galactic center so there is a large number of field stars visible in the near-IR image. In this field, 19 X-ray sources are identified by Rho et al. (2004) (black squares), while these and eight additional X-ray sources are found by MYStIX (red source extraction polygons). There are obvious UKIRT *K* band sources located at the positions of seven out of eight of these new sources. Two of the new X-ray sources are crowded with previously known X-ray sources, indicated by the reduced source-extraction polygons; both of these have *K*-band counterparts.

**NGC 1893** Figure 3 shows a portion of NGC 1893 imaged by *Chandra* and *Spitzer*. The star clusters in NGC 1893 form a band stretching from the north-east to the south-west of the ACIS-I field of view, and this portion of the field lies north of the south-west clusters. NGC 1893 is the most distant MSFR in the MYStIX sample at 3.6 kpc, but it also has one of the longest total exposure times of 446 ks. In this portion of the field, 28 X-ray sources were identified by Caramazza et al. (2012) (black squares), while these and 18 more were identified by MYStIX (red source-extraction polygons). At least seven of the new X-ray sources lie on mid-IR sources, although the lower resolution of *Spitzer* and the distance of the cluster mean that not all mid-IR counterparts are detected.

**RCW 38** Figure 4 shows the core of the cluster ionizing RCW 38. In this single *Chandra* exposure, the center of the cluster is slightly off-axis giving a characteristic four-lobed shape to the point-spread function shown by the red contours used for extraction. The core has a dense concentration of very bright massive and pre-main-sequence stars. In the field shown, the *wavdetect* procedure by Winston et al. (2011) located only 15 sources, missing many cases apparent to visual inspection. The MYStIX analysis gives 50 sources. Nearby sources are resolved with $\sim 1''$ separation; the inner $\sim 5''$ likely has more unresolved sources. We do not show an IR image in Figure 4 because there is no UKIRT observation, the 2MASS image lacks the spatial resolution to resolve IR counterparts in the core, and *Spitzer* images are overwhelmed by bright nebulosity.

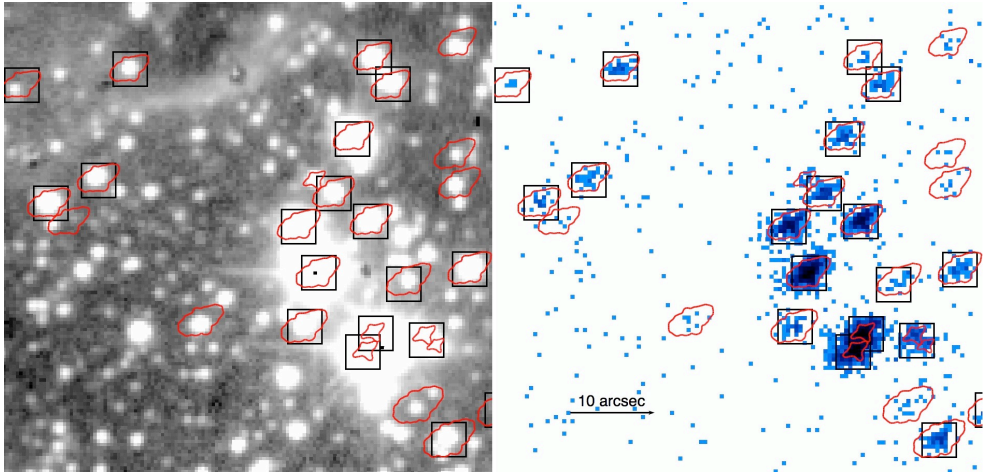

Fig. 2.— *Left:* UKIRT $K$-band image of a small region of the Trifid Nebula centered at $18^h\ 02^m\ 24\overset{s}{.}25\ -22°\ 01'\ 47''$ (J2000). *Right:* Chandra ACIS-I image of the same field of view shown here with $0.5''$ pixels. X-ray sources identified by Rho et al. (2004) are marked by black boxes, while MYStIX X-ray sources are indicated by the red polygonal extraction regions.

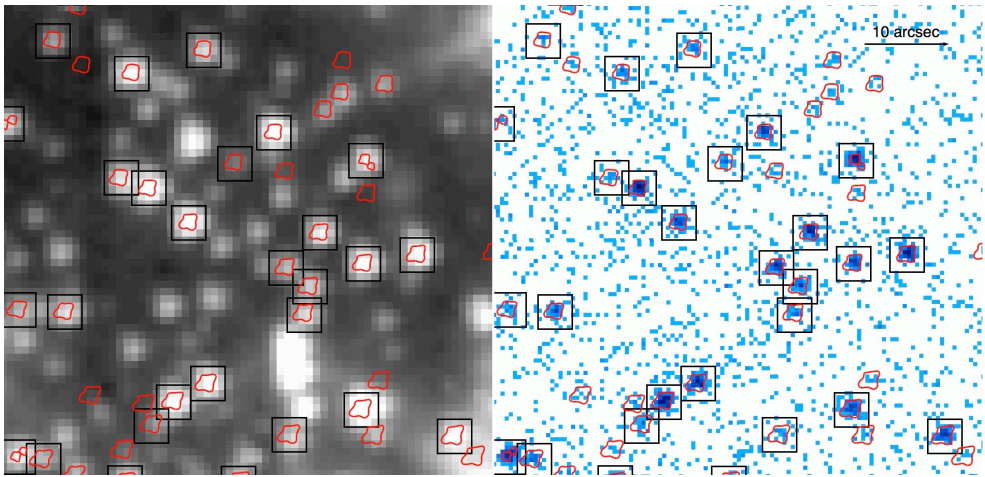

Fig. 3.— *Left:* Spitzer 3.6 $\mu$m-band image of a small region of NGC 1893 centered at $5^h\ 22^m\ 46\overset{s}{.}5\ +33°\ 26'\ 00''$ (J2000). *Right:* Chandra ACIS-I image of the same field of view shown here with $0.5''$ pixels. X-ray sources identified by Caramazza et al. (2012) are marked by black boxes, while MYStIX X-ray sources are indicated by the red polygonal extraction regions.

– 20 –

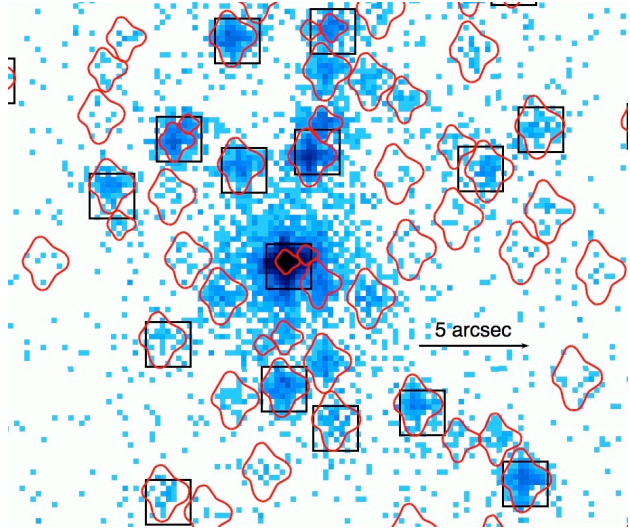

Fig. 4.— Center of the RCW 38 cluster observed with *Chandra* shown here with 0.25″ pixels. Red polygons show the MYStIX extraction regions, and boxes show the source locations of sources found with the *wavdetect* algorithm by Winston et al. (2011).

## 5. Contaminating X-ray Populations

Every *Chandra* exposure of a star-forming complex in the Galactic Plane will have three classes of unrelated X-ray sources, in addition to the young stellar population of interest: field stars in the foreground of the MSFR, more distant background stars, and extragalactic sources. The Monte Carlo simulations of their flux, median energy, and spatial distributions were performed using the procedures of Getman et al. (2011), with minor modifications described in Appendix A of Broos et al. (2013). Each MYStIX MSFR was treated individually with its unique mosaic of *Chandra* exposures, Galactic longitude and latitude, absorption for interstellar clouds, and distance to the complex. We use RCW 38 to illustrate the contamination here (Figure 5). The molecular clouds in this region are projected on the west side of the ACIS-I field of view, near the center of the young stellar cluster.

**Extragalactic sources** Mostly active galactic nuclei with some starburst galaxies, these are seen through the obscuring Galactic interstellar medium (Broos et al. 2007). A typical MYStIX exposure should have dozens of extragalactic sources, and the deeper mosaics should have hundreds. Realistic X-ray flux distributions (Moretti et al. 2003) and spectra (Brandt et al. 2001) of extragalactic sources were simulated, and were then subject to spatially-dependent absorption that was calculated using line-of-sight H<span>I</span> column density through the Galaxy (Dickey & Lockman 1990) and molecular cloud maps derived from CO observations or IR reddening of background stars (see Appendix A.2 in Broos et al. 2013). The simulated, absorbed sources were then superposed on the observed local background level and subject to the local MYStIX X-ray detection limit. The result of the simulation for RCW 38 is a prediction of $\sim 120$ extragalactic sources with a maximum source density of $\sim 0.8$ sources $(\text{arcmin})^{-2}$. Note from Figure 5 (left panel) that the extragalactic source spatial distribution is expected to be concentrated eastward of the *Chandra* field center, as the molecular cloud should obscure extragalactic sources westward of the field center. The contaminating source density is also higher toward the center of the field where the sensitivity to faint sources is higher.

**Foreground Galactic sources** Magnetically active, Galactic field stars, mostly younger main sequence stars, populate the line-of-sight towards RCW 38 at $d = 1.7$ kpc. These mostly have soft X-ray spectra and do not show spatial variations due to absorption by the molecular clouds in the star forming complex. The Monte Carlo calculation of their population is based on the Besançon Galactic structure model (Robin et al. 2003) and *ROSAT* surveys establishing X-ray luminosity functions as a function of stellar mass and age (Schmitt et al. 1995; Schmitt 1997; Hünsch et al. 1999). The simulation for the MYStIX RCW 38 observation predicts ∼ 60 foreground stellar X-ray sources with peak surface density around 0.4 sources (arcmin)$^{-2}$. These are assumed to be distributed uniformly across the sky, but, due to the higher sensitivity on-axis and intersecting CCD chip gaps in the ACIS-I detector, the expected spatial distribution of foreground stars shows a distinct pattern (Figure 5, right panel).

**Background Galactic sources** Field stars in the distant Galaxy are modeled including molecular cloud absorption similar to the extragalactic sources. Active and quiescent accretion binary systems are not included in the model. The simulation predicts ∼ 30 background X-ray sources in the MYStIX X-ray catalog for RCW 38 with a peak surface density around 0.2 sources (arcmin)$^{-2}$ (Figure 5, center panel). While always fewer than the extragalactic contaminants, the predicted background Galactic source population depends strongly on Galactic latitude and longitude.

Thus, the simulation for the RCW 38 *Chandra* field predicts ∼ 210 contaminating X-ray sources dominated by extragalactic sources. The contaminants typically have very little effect in the cores of rich stellar clusters, but can be a significant fraction of X-ray sources in peripheral regions of the *Chandra* fields. For example, in the northern corner of the RCW 38 field, about 1/3 of the X-ray sources are predicted to be contaminants.

These spatial maps of predicted contaminants, as well as their predicted X-ray median energies and counterpart *J* magnitudes, are used in the "naive Bayes" classifier described by Broos et al. (2013). Table 8 of that paper gives the predicted numbers of each class of contaminants for each MYStIX star forming complex. Generally, there is good agreement between the total number of predicted contaminants and the number of X-ray sources classified as contaminants or as "unclassified" due to insufficient information: for example, in RCW 38 the simulations predict 212 contaminants and the classifier identifies 206 sources as contaminants or "unclassified." The classifier also identifies 813 *Chandra* sources as probable young stars in the RCW 38 complex. Thus, 20% of the X-ray sources in this field are likely contaminants and 80% are likely scientifically interesting young stars. The agreement between contamination predictions and classifier results across the MYStIX sample (see Table 8 of Broos et al. 2013) gives confidence that the predictions are reasonably reliable.

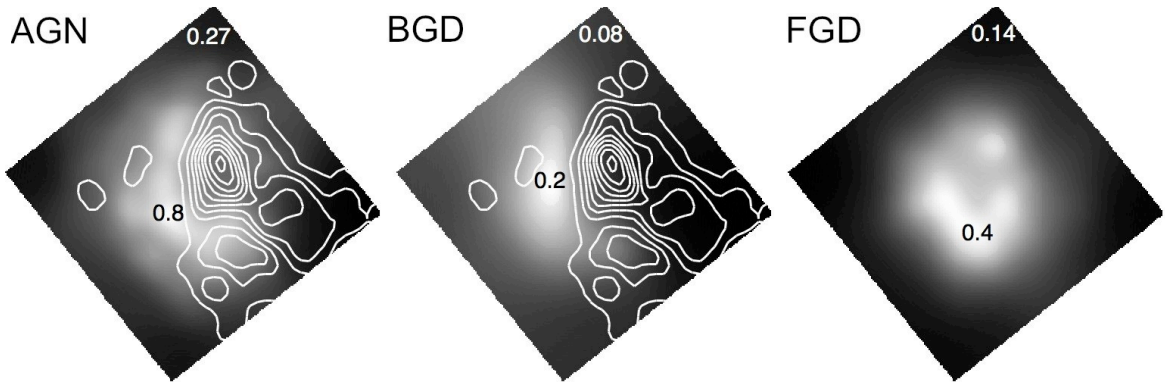

Fig. 5.— The predicted contaminating X-ray source populations in the MYStIX RCW 38 *Chandra* observation is shown using gray scale to represent the surface density of contaminants. The white contours represent a map of visual absorption calculated from $JHK$ field star distributions. The numbers in black and white give the peak and minimum values, respectively, in the surface density of predicted contaminants per square arcmin. *Left:* Extragalatic sources. *Center:* Background Galactic sources. *Right:* Foreground Galactic sources.

## 6. X-ray "Color−Magnitude Diagrams"

Figure 6 shows the X-ray "color−magnitude diagram" (median energy, $MedianEnergy\_t$, vs. incident energy flux in the total band, $F\_T$) for each of the ten regions. Sources have been color coded based on their classification as young stellar members (MPCMs), foreground stellar contaminants, background stellar contaminants, extragalactic contaminants, or unclassified sources by Broos et al. (2013). The distributions show that $F\_T$ varies over more than three decades, while $MedianEnergy\_t$ varies from <1 to ∼ 7 keV due to absorption. Absorption by interstellar gas increases $MedianEnergy\_t$ and decreases $F\_T$, shifting sources to the lower right.

Several effects can be seen in the distributions of the X-ray sources. Because detection is based on photon flux (photon/cm$^2$/s) rather than energy flux (erg/cm$^2$/s), an upward slope to the sensitivity limit as a function of $MedianEnergy\_t$ is seen in the lower part of the diagram. Most of the "unclassified" X-ray sources from Broos et al. (2013) are located near this detection limit. The MPCMs are mostly concentrated with $MedianEnergy\_t$ between $1 - 2$ keV, but extend to 7 keV due to absorption. The population of young stars with low absorption ($MedianEnergy\_t < 2$ keV) has a slight rise in $MedianEnergy\_t$ as flux increases, which is due to the increase in spectral hardness among more magnetically active young stars (e.g., Preibisch et al. 2005; Getman et al. 2010). In some regions such as Flame, RCW 36, DR 21, and RCW 38, almost all young stars are absorbed so there is a much broader distribution of MPCM $MedianEnergy\_t$ values. There is also a downward slope to maximum $F\_T$ as $MedianEnergy\_t$ increases. Young magnetically active stars typically are not seen with luminosities above $L_t \sim 10^{32}$ erg s$^{-1}$ (e.g., Getman et al. 2008), and the paucity of sources with simultaneous high flux and high $MedianEnergy\_t$ in the upper right of the diagram may be attributed to the effect of absorption, which shifts sources down and to the right on the plot.

The distribution of extragalactic sources on this diagram is shifted to the right of the lightly-absorbed MPCMs, but overlaps the distribution of highly-absorbed MPCMs. Extragalactic source median energies range from $2 < MedianEnergy\_t < 5$ keV. X-ray sources with $MedianEnergy\_t > 5$ are not classified as extragalactic by Broos et al. (2013) because the simulated extragalactic sources used for classification only include sources with photon indices $\Gamma > -0.5$ (Getman et al. 2011). Instead, the high-$MedianEnergy\_t$ sources are often classified as MPCMs by default, and some of these are likely to be protostars (cf. Getman et al. 2007).

The distributions of foreground and background stars on the diagram overlap with the distributions of the lightly-absorbed MPCMs, but their median energies are either low (for foreground stars) or high (for background stars) compared to most MPCMs. Median energy does help distinguish these contaminants; however, there is a relatively small number of these contaminants, and it is rare that the analysis by Broos et al. (2013) decisively indicates that an X-ray source is a field star. Rather, many field stars are left "unclassified" or occasionally misclassified.

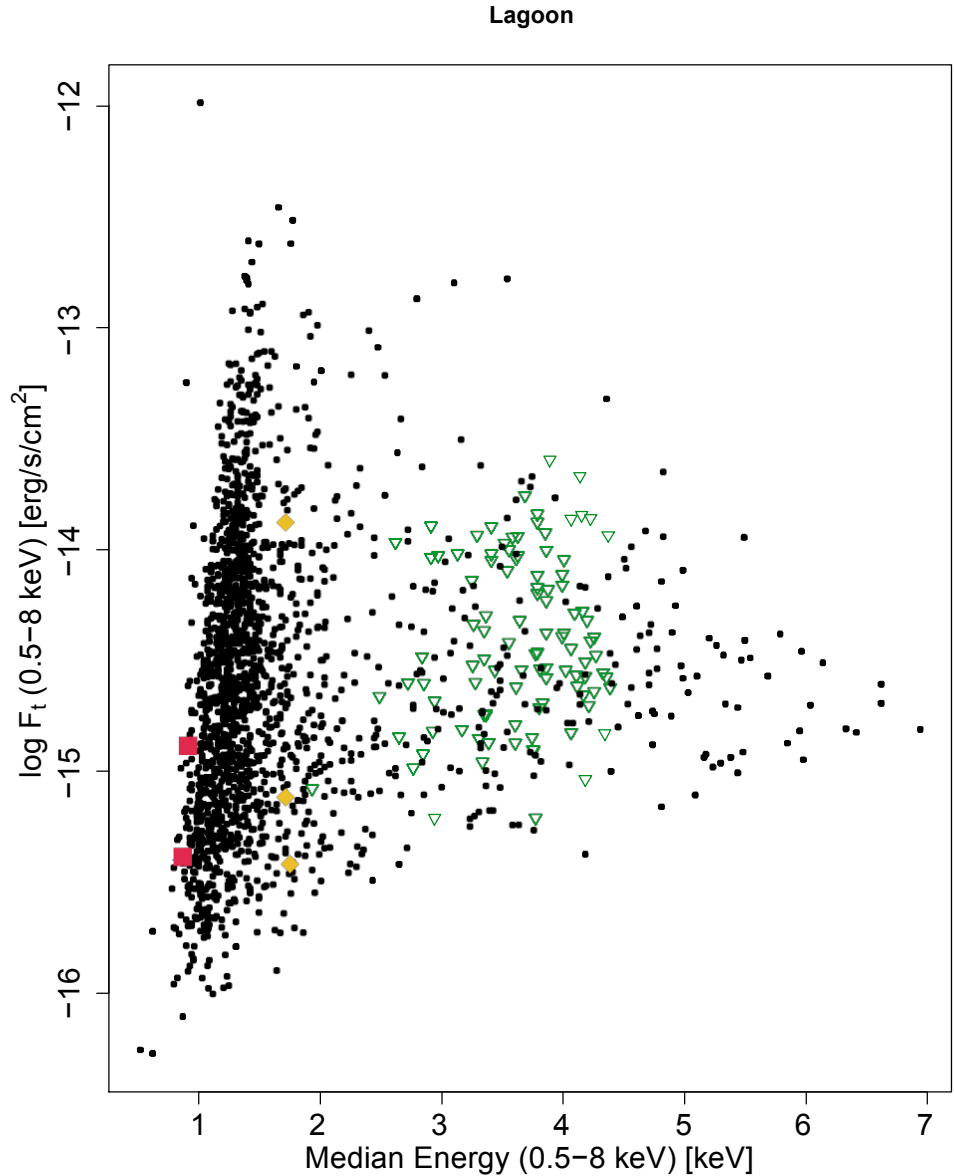

Fig. 6.— The X-ray "color−magnitude diagram" of the Lagoon Nebula with total-band flux (0.5 − 8.0 keV) vs. total-band median energy. X-ray sources are classified by Broos et al. (2013) as cluster members (black circles), foreground stars (red squares), background stars (orange diamonds), extragalactic sources (green triangles), and unclassified (blue crosses). Absorption by gas shifts sources to the lower right on this plot. Similar diagrams for the other nine MYStIX target regions are provided in the electronic version of this article.

Flame, RCW 36, NGC 2264, Rosette

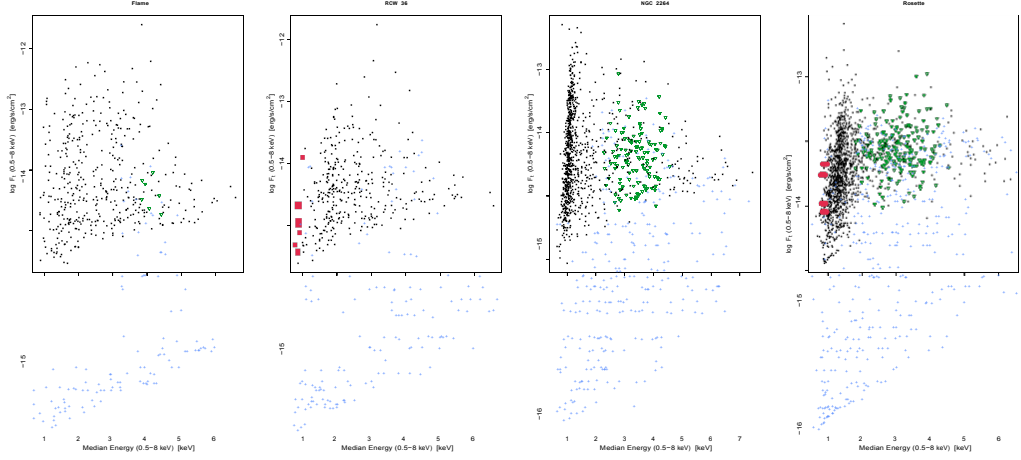

Lagoon, NGC 2362, DR 21, RCW 38

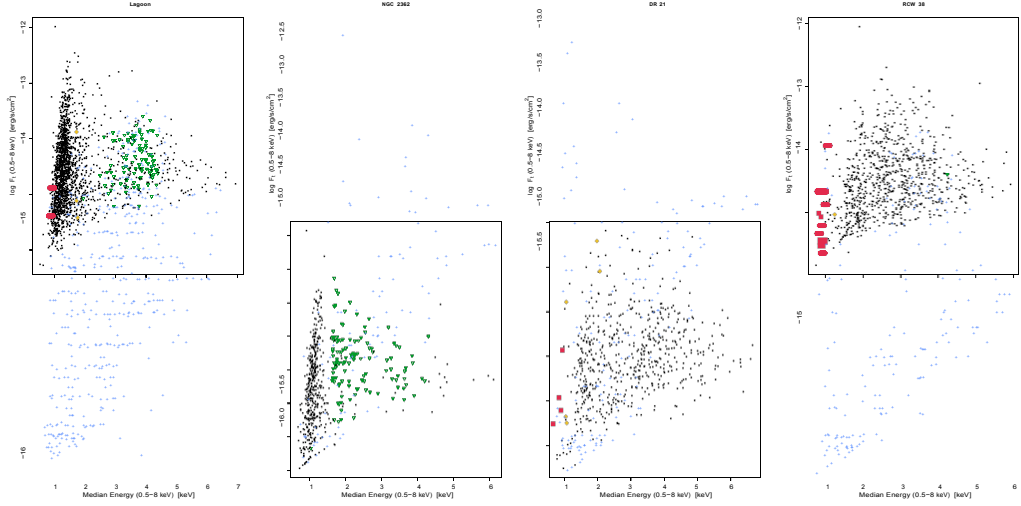

Trifid, NGC 1893

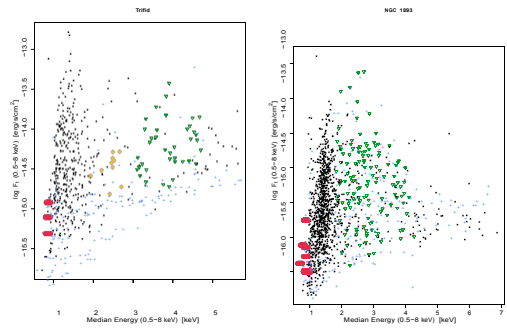

Fig. 6.— electronic figure set

## 7. Summary

We have described the identification of 11,315 faint X-ray sources in *Chandra* imaging observations of 10 massive star forming regions in support of the MYStIX project. Together with the X-ray sources reported in by Townsley & Broos (in preparation) and previously published for three regions (Orion Nebula, W 40, and the Carina complex), they provide an empirical foundation to the MYStIX effort to combine X-ray, near-IR, mid-IR, and published OB stellar catalogs to form multiwavelegth samples of young stars that range from the youngest protostars to older, disk-free, pre-main-sequence stars. The X-ray source lists will also be used to excise point-source X-rays from the *Chandra* data to reveal diffuse X-ray emission from the winds and supernovae of massive stars in these regions.

The data analysis is designed to be highly sensitive with sources as faint as 3 counts. Candidate sources are found from maximum likelihood reconstruction of the *Chandra* images that account for the spatially variable point-spread function, and the candidates are listed as confirmed sources when they satisfy a probabilistic threshold based on the local background. Source events are extracted using local point-spread functions and are characterized using nonparametric techniques (Broos et al. 2010; Getman et al. 2010). The procedures are closely based on those used in the recent *Chandra Carina Complex Project* (Townsley et al. 2011). Codes for many of these methods are publicly available[8].

Our X-ray catalogs are thus considerably larger than most previously published catalogs from the same datasets, allowing more X-ray sources to be associated with IR sources. The X-ray analysis is thus crucial for the production of rich samples classified as MPCMs (Broos et al. 2013). These MPCM samples are then used in MYStIX science studies such as identification and study of subclustering (Kuhn et al. 2013), derivation of X-ray/IR age estimators (Getman et al. 2013), and many other future planned studies. On the X-ray "color−magnitude diagram" the locus of lightly-absorbed MPCMs can be identified for some regions; however, it is difficult to distinguish between these sources and Galactic stellar contaminants or between highly-absorbed MPCMs and extragalactic contaminants from the X-ray data alone.

Acknowledgments:

The MYStIX project is supported at Penn State by NASA grant NNX09AC74G, NSF grant AST-0908038, and the *Chandra* ACIS Team contract SV4-74018 (G. Garmire &

---

[8]ACIS Extract (Broos et al. 2012) and XPHOT (Getman et al. 2012) are provided through the Astrophysics Source Code Library.

---